\documentclass[aps,pra,twocolumn,showpacs,superscriptaddress,floatfix]{revtex4-1}
\usepackage{graphicx,amsmath,amssymb}
\usepackage[usenames]{color}
\usepackage[dvipsnames]{xcolor}
\usepackage[colorlinks=true,linkcolor=blue,anchorcolor=blue,citecolor=blue,urlcolor=blue]{hyperref}

\usepackage{mathrsfs}
\usepackage{braket}
\usepackage{color}
\usepackage{dcolumn}% Align table columns on decimal point
\usepackage{bm}% bold math
\usepackage{flushend}  % "Figure X:" ¼Ó´Ö

\begin{document}

\title{
Phonon-induced two-axis spin squeezing with decoherence reduction in hybrid spin-optomechanical system
}

\author{Feng Qiao}
\affiliation{School of Physical Science and Technology, Lanzhou University, Lanzhou 730000, China}
\affiliation{Key Laboratory for Quantum Theory and Applications of MoE, Lanzhou Center for Theoretical Physics, Lanzhou University, Lanzhou 730000, China}

\author{Zu-Jian Ying}
\email{yingzj@lzu.edu.cn}
\affiliation{School of Physical Science and Technology, Lanzhou University, Lanzhou 730000, China}
\affiliation{Key Laboratory for Quantum Theory and Applications of MoE, Lanzhou Center for Theoretical Physics, Lanzhou University, Lanzhou 730000, China}

\begin{abstract}
We propose a scheme to implement Heisenberg-limited spin squeezing in a hybrid cavity optomechanical-spin system. In our system,
$N$ two-level systems are coupled via Tavis-Cummings interactions to a mechanical resonator (MR) in a standard
optomechanical setup. Within the dispersive coupling regime, adiabatic elimination of the optical mode induces a squeezing effect
on the MR, which, in the squeezed representation, effectively transforms the collective spin operators into a Bogoliubov form.
Under large detuning conditions, the phonon mode mediates interactions among the Bogoliubov collective spins, thereby enabling
a two-axis twisting squeezing protocol through appropriate parameter tuning. Both theoretical
analysis and numerical simulations show that in the presence of dephasing and phonon dissipation, the maximum squeezing degree
asymptotically converges to a constant as $N$ increases, which implies the metrological precision asymptotically approaches the standard quantum limit without parameter optimization. Nevertheless, in parameter optimization we extract a scaling relation of the optimal squeezing which surpasses existing schemes in the literature. Moreover, the optimization also leads to a considerable reduction of the preparation time for the optimal squeezing.
Our work may provide insights into dissipation effects in spin squeezing
and offer a potential route for high-precision quantum metrology in many-body systems.
\end{abstract}
\pacs{ }
\maketitle

%\date{\today}

\section{INTRODUCTION}\label{sec:1}

As a rapidly developing branch of quantum technology, quantum metrology exploits the nonclassical properties of various quantum states to achieve high-precision measurements~\cite{Degen2017-QuantSensing,Pezze2018entangelment,RamsPRX2018,PhysRevLett.96.010401,
Garbe2020,Montenegro2021-Metrology,Chu2021-Metrology,Garbe2021-Metrology,Ilias2022-Metrology,
Maccone2020Squeezing,Lawrie2019Squeezing,Gietka2023PRL2-Squeezing,Gietka2023PRL-Squeezing,Candeloro2021-Squeezing,
Ying2022-Metrology,Hotter2024-Metrology,Alushi2024PRL,Mukhopadhyay2024PRL,Mihailescuy2024,Ying-Topo-JC-nonHermitian-Fisher,
Ying-g2hz-QFI-2024,Ying-gC-by-QFI-2024,Ying-g1g2hz-QFI-2025,Ying2025g2A4,Ying2025TwoPhotonStark}.
Both theoretically and experimentally squeezing~\cite{Maccone2020Squeezing,Lawrie2019Squeezing,Gietka2023PRL2-Squeezing,Gietka2023PRL-Squeezing,Candeloro2021-Squeezing,
Ying2015,Ying-g1g2hz-QFI-2025,Ying2025g2A4,Ying2025TwoPhotonStark,Ying-gapped-top,
GaoXL2024PRAsquz,BaiSY2021PRL,LiPB2024AQTspinSquz,LiPB2021PRASpinSquz,WuLA2024PRAenvironSquz,KuangLM2024squz} is one of the most extensively studied quantum resources for quantum metrology~\cite{Maccone2020Squeezing,Lawrie2019Squeezing,Gietka2023PRL2-Squeezing,Gietka2023PRL-Squeezing,Candeloro2021-Squeezing,
Ying-g1g2hz-QFI-2025,Ying2025g2A4,Ying2025TwoPhotonStark,Ying-gapped-top}, alongside with quantum entanglement~\cite{Pezze2018entangelment} and critical
behavior of quantum phase transitions~\cite{Garbe2020,Montenegro2021-Metrology,Chu2021-Metrology,Garbe2021-Metrology,Ilias2022-Metrology,Ying2022-Metrology,
Hotter2024-Metrology,Alushi2024PRL,Mukhopadhyay2024PRL,Mihailescuy2024,Ying-Topo-JC-nonHermitian-Fisher,Ying-g2hz-QFI-2024,Ying-gC-by-QFI-2024,Ying-g1g2hz-QFI-2025,Ying2025g2A4,Ying2025TwoPhotonStark}.

Notable applications of squeezing include gravitational-wave detection and axion dark matter searches~\cite{goda2008quantum,aasi2013enhanced,yu2020quantum,backes2021quantum}, more accurate atomic clocks~\cite{PhysRevLett.104.250801,komar2014quantum,pedrozo2020entanglement},  high-precision quantum gyroscopes and magnetometers~\cite{PhysRevLett.114.063002,PhysRevLett.109.253605,PhysRevLett.113.103004}, and so forth.
In this regard, apart from traditional light squeezing~\cite{goda2008quantum,aasi2013enhanced,yu2020quantum,backes2021quantum,PhysRevLett.114.063002,LiJieYouJQ2022SqueezingAgaistTemperature}, spin squeezing has been arising as a novel squeezing resource~\cite{PhysRevLett.104.250801,komar2014quantum,pedrozo2020entanglement,PhysRevLett.109.253605,PhysRevLett.113.103004,
BaiSY2021PRL,LiPB2024AQTspinSquz,LiPB2021PRASpinSquz}. It leverages pairwise entanglement among spins to establish quantum correlations~\cite{PhysRevA.47.5138,wang2002pairwise,PhysRevA.68.012101} that circumvent the classical precision bound --- the shot-noise limit. Specifically, when initially prepared coherent spin state (CSS) (an product state of uncorrelated spins) is driven by various interaction mechanisms,
projection noise is redistributed
from the binomial distribution to a sub-binomial distribution characteristic of spin-squeezed state (SSS)~\cite{ma2011quantum}.
Schemes to generate spin squeezing primarily include photon-spin interactions~\cite{wineland1994squeezed,RevModPhys.82.1041,zhang2017cavity,qin2020strong},
quantum nondemolition measurements~\cite{kuzmich1998atomic,kuzmich2000generation,kritsotakis2021spin}, and the widely adopted one-axis twisting (OAT) and two-axis twisting (TAT) protocols~\cite{PhysRevA.47.5138,PhysRevA.91.043642,bohnet2016quantum,hosten2016quantum,hosten2016measurement}. In particular, the TAT scheme is recognized for achieving Heisenberg-limited spin scaling of $1/N$. However, its practical implementation generally requires complex driving laser fields or continuous control over the atomic ensemble~\cite{PhysRevA.80.032311,PhysRevLett.107.013601,PhysRevLett.132.113402}.

On the other hand, cavity optomechanics may open an alternative avenue for implementation of spin squeezing. Cavity optomechanics investigates the interaction between electromagnetic field with nano- and micro-mechanical resonator (MR), mediated by the radiation pressure of the confined optical field~\cite{RevModPhys.86.1391,xiong2015review}. Recently, experimental advances have even achieved coupling with chip-scale oscillators~\cite{stettenheim2010macroscopic,safavi2013squeezed,renninger2018bulk}. Within this platform, a wealth of theoretical and experimental progresses have been achieved, including optomechanically induced transparency~\cite{weis2010optomechanically,wang2014optomechanical,ma2014tunable}, ground-state cooling of mechanical motion~\cite{PhysRevLett.124.173601,whittle2021approaching,gan2019intracavity}, applications to quantum computation and communication~\cite{PhysRevLett.105.220501,safavi2011proposal,yan2015entanglement}, and the exploration of nonlinear dynamical phenomena such as period-doubling, optomechanical chaos~\cite{ma2014formation}. In such situations, several studies have examined squeezing of the optical and mechanical modes~\cite{nunnenkamp2010cooling,purdy2013strong,safavi2013squeezed}, and hybrid coupling of atomic ensembles and MR have enabled photon-triggered phase transitions and the transfer of squeezing between phonon mode and spin ensemble~\cite{lu2018single,zhang2021single}. In such hybrid optomechanical systems, effort to realize TAT relies on complex OAT interaction control~\cite{xie2024unitaryefficientspinsqueezing}. Achieving Heisenberg-limit spin squeezing with high efficiency remains to be a challenge, and the asymptotic behavior of optimal squeezing under realistic dissipative dynamics has yet to be fully characterized.

In the present work, we demonstrate a versatile Heisenberg-limited spin squeezing protocol in a hybrid system comprising $N$ two-level systems (representing spins) and an optomechanical platform. In our configuration, the MR is simultaneously coupled to both optical cavity mode and atomic ensemble. Upon linearization of the optomechanical interaction~\cite{PhysRevLett.98.030405} and operating in the regime where the frequencies of atomic ensemble and optical mode are far detuned from the phonon frequency, the photonic and phononic degrees of freedom can be eliminated. The resulting effective Hamiltonian reduces to a Bogoliubov-type collective spin-spin interaction,
which directly yields a tunable interpolation between OAT and TAT dynamics, thereby realizing a versatile spin-squeezing scheme.

Considering two-body losses in a two-mode Bose-Einstein condensate, the asymptotic optimal squeezing of OAT protocol in the limit $N \rightarrow \infty$ is independent of $N$~\cite{refId0,PhysRevLett.100.210401}. An analogous result is observed in our system, when individual spin dephasing and mechanical dissipation are incorporated in the TAT protocol, we find that the optimal squeezing parameter asymptotically approaches a constant determined by both intrinsic system parameters and dissipation rates. In the ideal situation, this constant vanishes, recovering the Heisenberg scaling. By applying a standard linearization approach in the large-$N$ limit, we derive an explicit expression for this asymptotic value. Our analysis reveals that, under the influence of Markovian noise, the metrological precision enabled by spin squeezing asymptotically returns to the standard quantum limit as $N$ increases, in agreement with the conclusion of Ref.~\cite{Fujiwara_2008}. Furthermore, with parameter optimization we finally extract a scaling relation of the optimal squeezing parameters capable of surpassing the existing schemes in the literature. In addition, the optimization also leads to reduction of preparation time for the optimal squeezing. These findings may have significant implications for the theoretical and experimental pursuit of spin squeezing in realistic open quantum systems.

The paper is organized as follows.
Sec.~\ref{sec:2} introduces the hybrid optomechanical model and the effective Hamiltonian is derived.
Sec.~\ref{sec:3} addresses the phonon-mediated collective spin-spin interaction which enables diverse spin-squeezing protocols.
Sec.~\ref{sec:dynmics-no-dissipation} illustrates the dynamics for the squeezing parameter in the absence of dissipation.
Sec.~\ref{sec:dissipation} investigates the dynamics of the squeezing parameter in the presence of dissipation with the following results: (i) The asymptotic behavior of the minimum squeezing parameter is revealed for fixed system parameters in the large-spin-number limit; (ii) The improved scaling relation for optimal squeezing parameter is extracted in the optimization of system parameters; (iii) The time to reach optimal squeezing is also reduced in optimization.
Finally, Sec.~\ref{sec:CONCLUSION} provides a summary of our results.

\section{MODEL AND HAMILTONIAN}\label{sec:2}

\begin{figure}[t]
	\centering
  \includegraphics[width=1.0\columnwidth]{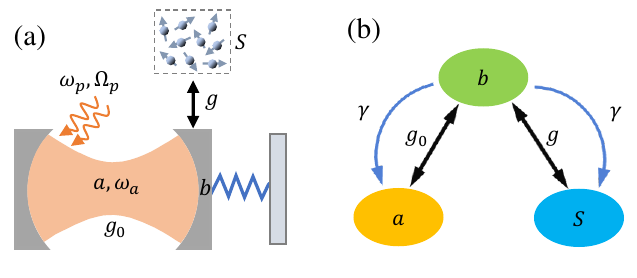}.
	\caption{Scheme of the hybrid optomechanical platform for Heisenberg-limited spin squeezing.
  (a) A mechanical resonator (MR) with phonon mode $b$ is coupled to an optical cavity mode $a$ with single-photon optomechanical coupling $g_{0}$, and simultaneously interacts with an atomic ensemble $S$ (spins in square) with coupling strength $g$. The cavity is driven by an external laser of frequency $\omega_{p}$ and amplitude $\Omega_{p}$.
  (b) Phonon dissipation at rate $\gamma$ induces effective dissipative channels on both the optical and spin degrees of freedom.
  }
  \label{fig1}
\end{figure}

We consider a hybrid optomechanical system consisting of an atomic ensemble coupled to a MR, as schematically depicted in Fig.~\ref{fig1}. Here, as sketched in Fig.~\ref{fig1}(a), the MR labelled by $b$ may be a movable mirror in Fabry-P\'{e}rot cavity~\cite{RevModPhys.86.1391}, a vibrational nanostructure~\cite{PhysRevLett.110.156402} or other mechanical oscillators~\cite{chan2011laser,riedinger2016non}. The MR can be coupled to the cavity mode (labelled by $a$) via radiation pressure~\cite{chan2011laser}. Simultaneously, the movable mirror interacts with the atomic ensemble (labelled by $S$) via a standard Tavis-Cummings (TC)
interaction~\cite{PhysRevLett.117.210503}. Additionally, the optical cavity is externally driven by classical fields. As a consideration for more realistic situations in open quantum systems phonon dissipation at rate $\gamma$ will be further introduced, which induces effective dissipative
channels on both the optical and spin degrees of freedom, as indicated in Fig.~\ref{fig1}(b).

In the rotating frame of the driving field, the system Hamiltonian is given by ($\hbar = 1$)
\begin{align}\label{con:1}
  \hat{H}&=\hat{H}_{\rm TC}+\hat{H}_{om}+\hat{H}_{dr},
\end{align}
with
\begin{subequations}\label{con:2}
  \begin{align}
    \hat{H}_{om}&=\Delta _a\hat{a}^{\dagger}\hat{a}+g_0\hat{a}^{\dagger}\hat{a}(\hat{b}+\hat{b}^{\dagger}),\\
    \hat{H}_{\rm TC}&=\omega _b\hat{b}^{\dagger}\hat{b}+\Omega \hat{S}_z+g(\hat{b}\hat{S}_++\hat{b}^{\dagger}\hat{S}_-),\\
	  \hat{H}_{dr}&=\Omega _p(\hat{a}^{\dagger}+\hat{a}),
  \end{align}
\end{subequations}
The Hamiltonian $\hat{H}_{om}$ consists of two terms: the first term represents the free Hamiltonian of the optical field in the quantized cavity mode $a$, while the second term describes the optomechanical coupling with a coupling strength $g_0$. Here, $\hat{a}(\hat{a}^{\dagger})$ and $\hat{b}(\hat{b}^{\dagger})$ denote the annihilation (creation) operators of the optical field and the MR with phonon mode $b$, respectively. The parameter $\Delta_a = \omega_a - \omega_p$ gives the detuning frequency of cavity $a$ (with frequency $\omega_a$) relative to the external driving laser (with frequency $\omega_p$). The term $\hat{H}_{\rm TC}$ corresponds to the TC model Hamiltonian, which describes the interaction between the MR and the atomic ensemble $S$. The coupling strength is denoted by $g$, while $\omega_b$ and $\Omega$ represent the resonant frequency of the phonon mode and the transition frequency of the atomic ensemble, respectively. The collective spin operator is given by $\hat{S}_k=\frac{1}{2}\Sigma _i\hat{\sigma}_{k}^{(i)},k\in \{x,y,z\}$, where $\hat{\sigma}_{k}^{(i)}$ denotes the Pauli operator acting on the $i$-th spin, and $\hat{S}_{\pm}=\hat{S}_x+i\hat{S}_y$. The term $\hat{H}_{dr}$ represents the external driving laser field, characterized by a driving amplitude $\Omega_p$ and frequency $\omega_p$.

Under strong laser driving, the cavity mode $a$ and phonon mode $b$ can be expressed as their mean-field values plus quantum fluctuation operators around these averages, i.e., $\hat{a} = \alpha+ \delta \hat{a} $ and $\hat{b}=\beta+ \delta \hat{b} $, where $\alpha$ and $\beta$ denote the expectation values of the two annihilation operators, while
$\delta \hat{a} $ and $\delta \hat{b}$ represent the fluctuation operators. Substituting these expressions into the Hamiltonian and neglecting higher-order fluctuation terms, the nonlinear optomechanical interaction can be linearized. For simplicity, we retain the notations $ (\hat{a}, \hat{b}) $ to represent the fluctuation operators $ (\delta \hat{a}, \delta \hat{b}) $. The resulting effective linearized Hamiltonian is then derived as (see the derivation in Appendix~\ref{Appendix-Linearize-OM})
\begin{align}\label{con:3}
  \hat{H}_L&=\Delta\hat{a}^\dagger\hat{a}+\omega_b\hat{b}^\dagger\hat{b}+G\left(\hat{a}+\hat{a}^\dagger\right)\left(\hat{b}+\hat{b}^\dagger\right)\nonumber
\\
  &+\Omega\hat{S}_z+g\left(\hat{b}\hat{S}_++\hat{b}^\dagger\hat{S}_-\right),
\end{align}
Due to the linearization procedure, the cavity-field detuning frequency relative to the driving field shifts to $ \Delta=\Delta_a + 2g_0\beta $, while the optomechanical coupling strength is enhanced to $G=g_0\alpha $. For simplicity, we select an appropriate driving phase such that the expectation values $\alpha$ and $\beta$ remain real.

Next, we consider the regime of large detuning between the cavity field frequency $\Delta $ and the MR frequency $\omega_b$, where the optomechanical coupling strength is much smaller than the detuning frequency, i.e., $G\ll \left| \Delta -\omega _b \right| $. In this regime, there is no significant energy exchange between the optical field and the MR. To eliminate the cavity mode $a$, we perform a unitary transformation $U = e^{\hat{V}}$ on the Hamiltonian $\hat{H}_L$, where the generator is given by
\begin{align}\label{con:4}
  \hat{V}=\mu \left(\hat{a}\hat{b}-\hat{a}^{\dagger}\hat{b}^{\dagger}\right) +\nu \left(\hat{a}\hat{b}^{\dagger}-\hat{a}^{\dagger}\hat{b}\right),
\end{align}
with $\mu =G/\left( \Delta +\omega _b \right)$ and $\nu =G/\left( \Delta -\omega _b \right)$. Keeping terms up to the second order, the resulting effective Hamiltonian is given by (see the derivation in Appendix~\ref{Appendix-Heff})
\begin{align}\label{con:5}
  \hat{H}_{eff}=\omega_b\hat{b}^\dagger\hat{b}+\Omega\hat{S}_z+g\left(\hat{b}\hat{S}_++\hat{b}^\dagger\hat{S}_-\right)-\Gamma\left(\hat{b}+\hat{b}^\dagger\right)^2.
\end{align}
Here the last term represents the nonlinear contribution of the photon-induced phonon mode $b$, with a coefficient given by
\begin{equation}
\Gamma =\Delta G^2/(\Delta ^2-{\omega _b}^2).
\end{equation}
By applying the squeezing transformation $\hat{U}(r) =\exp[r( \hat{b}^2-\hat{b}^{\dagger 2}) /2]$ to the effective Hamiltonian in Eq.\eqref{con:5} and choosing $r=(1/4)\ln\mathrm{[}1-4\Gamma /\omega _b]$, the nonlinear term can be eliminated. We discard constant terms that do not affect the system's dynamics. Furthermore, for convenience, we define a reformed atomic ensemble mode operator as $\hat{\varXi}\equiv\hat{S}_-\cosh r-\hat{S}_+\sinh r$, the effective Hamiltonian is then simplified to
\begin{align}\label{con:6}
  \hat{H}_{eff}=\omega_{r}\hat{b}^\dagger\hat{b}+\Omega\hat{S}_z+g\left(\hat{b}\hat{\varXi}^\dagger+\hat{b}^\dagger\hat{\varXi}\right).
\end{align}
Thus, the resonance frequency of the phonon mode is modified by an exponential factor due to the squeezing transformation, given by $\omega_r = e^{2r} \omega_b$. Meanwhile, the effect on the atomic ensemble is equivalent to a rotation in the collective spin operator space.

\section{Phonon-Induced Spin-Spin Interactions and diverse spin-squeezing protocols}\label{sec:3}

\subsection{Phonon-induced spin-spin interactions}

In the effective Hamiltonian Eq.~\eqref{con:6}, the phonon mode frequency $\omega_r$ is ultimately determined by the original phonon frequency $\omega_b$, the optomechanical coupling strength $g_0$, and the cavity detuning $\Delta$. Therefore, by appropriately choosing the system parameters, the renormalized mechanical frequency $\omega_r$ can be tuned to be in the dispersive (far-detuned) regime relatively to the atomic ensemble transition frequency, i.e., $|\omega_r - \Omega| \gg g $. In this regime, the virtual excitations of the mechanical mode mediate an effective interaction between the two-level atoms. Applying the standard Schrieffer-Wolff transformation $\hat{R}=\exp[ g/\omega_{r}\left( \hat{b}^{\dagger}\hat{\Sigma}-\hat{b}\hat{\Sigma}^{\dagger} \right)]$ and retaining terms up to second order, the Hamiltonian can be rewritten as
\begin{align}\label{con:7}
  \hat{H}_{eff}\simeq \omega_{r}\hat{b}^\dagger\hat{b}+\Omega\hat{S}_z-2\chi\hat{b}^\dagger\hat{b}\hat{S}_z-\chi\hat{\varXi}^\dagger\hat{\varXi}.
\end{align}
Here, $\chi = g^2/\omega_r $ denotes the coupling strength between the rotated spin ensemble operators $\hat{\varXi}$ induced by the mechanical mode. Next, by preparing the phonon mode in its ground state to eliminate the degrees of freedom associated with the MR in Eq.\eqref{con:7}, we expand and reorganize the two terms corresponding solely to the spin part as follows:
\begin{align}\label{con:8}
  \hat{H}_{s} =-\tilde{\chi}\left[\left(\hat{S}^{2}-\hat{S}_{z}^{2}\right)-\mathrm{tanh}2r\left(\hat{S}_{x}^{2}-\hat{S}_{y}^{2}\right)\right]+\tilde{\Omega}\hat{S}_{z}.
\end{align}
Here, we have defined the parameters $\tilde{\chi} \equiv \chi \cosh 2r$, $\tilde{\Omega} \equiv \Omega - \chi$, and $\hat{S}^2 = \hat{S}_{x}^{2} + \hat{S}_{y}^{2} + \hat{S}_{z}^{2}$ denotes the square of the total spin of the atomic ensemble.

\subsection{Diverse spin-squeezing protocols}

In the effective Hamiltonian Eq.~\eqref{con:8}, the first two terms constitute second-order nonlinear contributions, which are responsible for the spin squeezing effect in the ensemble. Since the last term does not induce squeezing, we disregard its influence on the system. Noting that
\begin{equation}
\tanh 2r = 2\Gamma/(2\Gamma - \omega_b),
\end{equation}
we can control the magnitude of $\Gamma$ by adjusting the system parameters to realize different squeezing schemes. For example, by setting $\Gamma = 0$ (i.e., $G=0$ or $\Delta=0$), we implement the OAT spin squeezing scheme~\cite{PhysRevA.47.5138}. Here, $G=0$ indicates the absence of optomechanical coupling in the system, reducing our approach to the form presented in Ref.~\cite{PhysRevLett.110.156402}. Moreover, $\Delta=0$ signifies that the effective frequency of the optical field in the optomechanical system, $\Delta$, is taken as the zero-frequency reference.

By setting $\Gamma = -\omega_b/4$ (where $\tanh2r = 1/3$), the parameter is modified to $\tilde{\chi} = 3\chi/2\sqrt{2}$, allowing the Hamiltonian to be rewritten as:
\begin{align}\label{con:9}
  \hat{H}_{s}  =-\frac{\sqrt{2}}{2}\chi(\hat{S}^{2}+\hat{S}_{y}^{2}-\hat{S}_{z}^{2}).
\end{align}
Since the total spin-squared operator $S^2$ is a conserved quantity, it does not participate in the dynamical evolution of the system. Consequently, the Hamiltonian is equivalent to a standard TAT squeezing scheme in the $y$-$z$ plane, which can achieve Heisenberg-limited spin squeezing. We emphasize that while our squeezing protocol aligns with Ref.~\cite{PhysRevLett.125.203601} in form, the control parameter $\Gamma$ in our system depends jointly on external driving lasers, photon frequencies, and phonon frequencies. This broader parameter space grants our scheme a significant experimental flexibility. Furthermore, adjusting $\Gamma =1/8\omega_b$ (corresponding to $\tanh 2r = -1/3$) realizes a TAT squeezing protocol in the $x$-$z$ plane. For generic values of $\Gamma$, the Hamiltonian corresponds to a weighted hybrid two-axis squeezing scheme. This tunability establishes our system as a versatile platform for exploring diverse spin-squeezing protocols.

\begin{figure}[t]
	\centering
	\includegraphics[width=1.0\columnwidth]{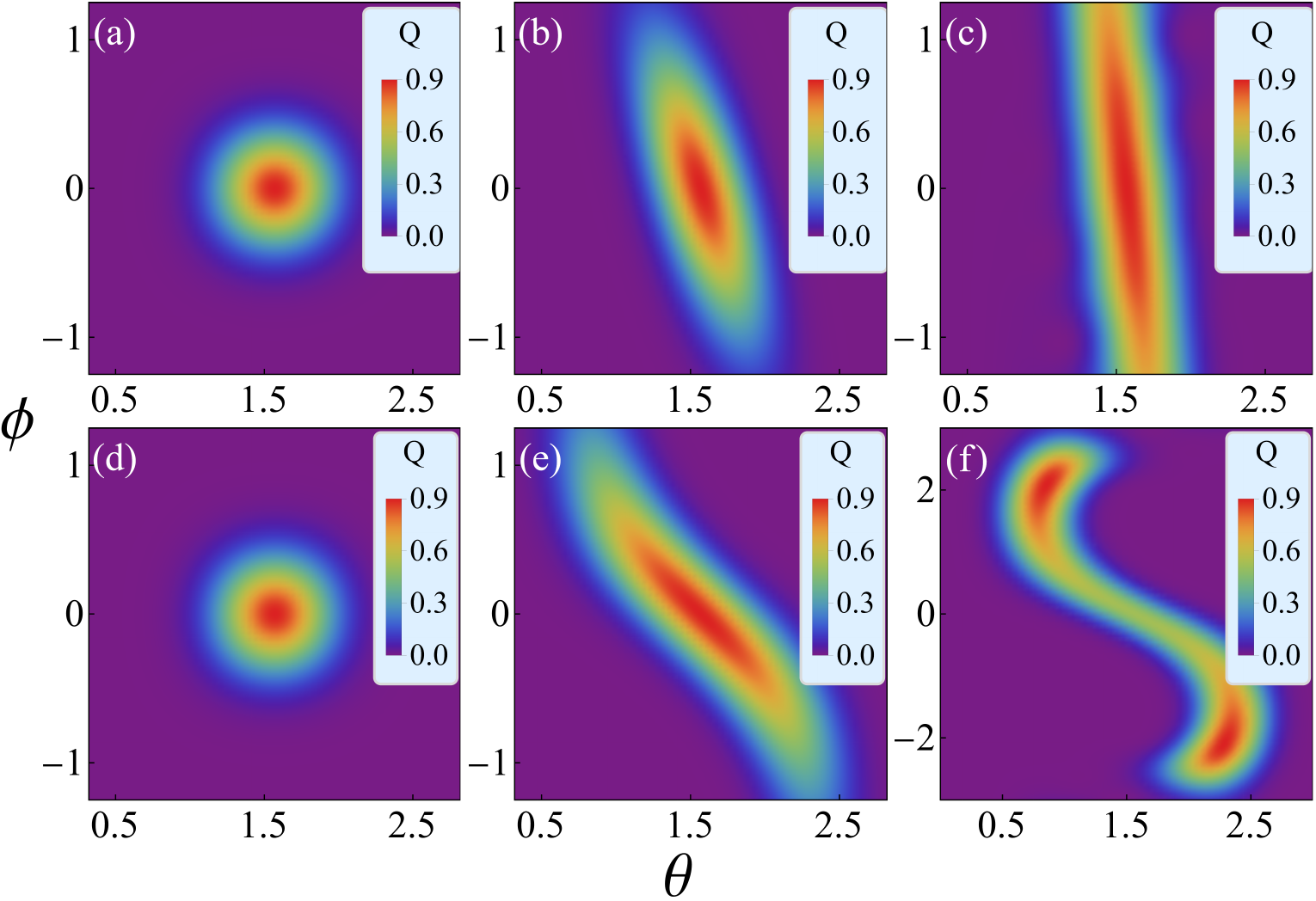}.
	\caption{The Husimi $Q$ function at different moments of time $t$ under the one-axis twisting (OAT) [(a)-(c)] and
two-axis twisting (TAT) [(d)-(f)] schemes. Here $\theta$ and $\phi$ denote the polar and azimuthal angles. Panels (a) and (d) correspond to the initial  coherent spin state (CSS) $|\mathrm{\pi/2,0}\rangle$ at $t=0$. Panels (b) at $t=0.105\chi^{-1}$ and (e) at $t=0.051\chi^{-1}$ show the quasiprobability distributions at the maximum-squeezing times. Panels (c) at $t=0.3\chi^{-1}$ and (f) at $t=0.1\chi^{-1}$ depict the oversqueezing.
         }
  \label{fig2}
\end{figure}

\subsection{Illustrations for spin squeezing by the Husimi Q function}

We initially prepare the system in a CSS with its collective spin aligned along the $x$-axis, denoted as $|\pi /2,0\rangle$. This state satisfies $\hat{S}_x|\pi /2,0\rangle =N/2|\pi /2,0\rangle$ and exhibits isotropic fluctuations in the $y$-$z$ plane, with $\Delta \hat{S}_y = \Delta \hat{S}_z = N/4$. Under the time evolution driven by the system Hamiltonian, the initial product state evolves into a many-body entangled state characterized by pairwise entanglement~\cite{PhysRevA.47.5138,wang2002pairwise}. Consequently, quantum fluctuations redistribute in the plane $\mathbf{n}_\perp$ perpendicular to the mean spin direction $\mathbf{n}$, with squeezing along one quadrature at the expense of anti-squeezing in the orthogonal direction. The spin state is defined as squeezed if the minimal variance falls below the standard quantum limit~\cite{PhysRevA.47.5138} of $N/4$ for a CSS, i.e., $\langle (\Delta S)^2 \rangle_{\text{min}} < N/4$.

To illustrate this squeezing effect, we present the evolution of the quasi-classical probability distribution described by the Husimi $Q$ function
\begin{equation}
Q\left( \theta ,\varphi \right) =\left( 2j+1\right) /\left( 4\pi \right)
\left\langle \theta ,\varphi \right\vert \hat{\rho }\left\vert \theta
,\varphi \right\rangle
\end{equation}%
under both the OAT and TAT schemes. We have defined
$\left\vert \theta ,\varphi \right\rangle
=( 1+\left\vert \eta \right\vert ^{2} )^{-j}e^{-\eta ^{\ast }\hat{S}^{-}}\left\vert j,j\right\rangle $
(here $j=N/2$), where $\left\vert j,m\right\rangle $ $(m=-j,\cdots ,j)$ is the
eigenstate of $\{ \hat{S}^{2},\hat{S}_{z} \} $, $\eta = - \tan \left( \theta /2\right) e^{-i\varphi }$.
The Husimi $Q$ function is a quasiprobability distribution of a quantum state in the phase space (here
defined by $\theta ,\varphi $). As evidently seen in Fig.~\ref{fig2}, the
distribution is progressively squeezed over time. Indeed, there is no squeezing in the initial CSS as shown by the round profiles of the
Husimi $Q$ function in Figs.~\ref{fig2}(a) (OAT) and \ref{fig2}(d) (TAT). With the time evolution the flattened profiles in Figs.~\ref{fig2}(b),(c),(e),(f) reflect the evident squeezing. Here Figs.~\ref{fig2}(b),(e) are the maximum squeezing cases while Figs.~\ref{fig2}(c),(f) are oversqueezing cases.

\section{Squeezing degree in different squeezing schemes in the absence of dissipation}\label{sec:dynmics-no-dissipation}

Although the squeezing can be qualitatively visualized by the pictorial representation in the Husimi $Q$ function shown in the last section, we need to analyze the squeezing degree quantitatively by the squeezing parameter. Here in this section we first consider the ideal situation in the absence of dissipation.

\subsection{Squeezing parameter for squeezing degree}

To quantify the degree of squeezing achieved by these different schemes, the Ramsey squeezing parameter $\xi _{R}^{2}$ is introduced~\cite{PhysRevA.50.67}:
\begin{align}\label{con:10}
  \xi _{R}^{2}=\frac{N \langle (\Delta S)^2 \rangle_{\perp}}{|\langle S\rangle|^2},
\end{align}
where $\langle (\Delta S)^2 \rangle_{\perp}=(V_+-\sqrt{V_{-}^{2}+4V_{yz}^{2}})/2$ denotes the minimum fluctuation in the $\mathbf{n}_\perp$ plane~\cite{ma2011quantum} described above. Here $V_{\pm}=\langle S_{y}^{2}\pm S_{z}^{2}\rangle$ and $V_{yz}=\langle S_yS_z+S_zS_y\rangle /2$, and $|\langle S \rangle|$ represents the magnitude of the total spin expectation value. Spin squeezing is achieved when $\xi _{R}^{2}<1$. A smaller value of $\xi _{R}^{2}$ will mean a larger degree of squeezing.

The squeezing parameter $\xi_{R}^{2}$ quantifies the ratio of phase fluctuations in a given state relative to CSS. A value of $\xi_{R}^{2} \geq 1$ indicates the absence of quantum metrological advantage, whereas $\xi_{R}^{2} < 1$ signifies the ability to surpass the standard quantum limit (SQL) in phase sensitivity. This parameter thus serves as a critical metric in quantum metrology for evaluating performance beyond classical benchmarks.

\subsection{Time evolution of the squeezing parameter for different squeezing schemes}

In the absence of dissipation, the dynamics is completely governed ed by the system Hamiltonian $H$ in the way that an initial state $\psi(0)$ will evolve by $\psi(t)=e^{-iHt}\psi(0)$. We first investigate the time evolution of the squeezing parameter for four distinct schemes, as shown in Fig.~\ref{fig-Squz-time-no-Dissipation}. The figure reveals that as the magnitude of $\Gamma$ increases, the squeezing scheme transits from OAT to TAT and then to two-axis schemes with different weightings. Correspondingly, the optimal squeezing time decreases, while the optimal squeezing degree reaches a maximum (as the squeezing parameter $\xi _{R}^{2}$ reaches a minimum) at TAT and subsequently decreases ($\xi _{R}^{2}$ increases). These results demonstrate the superiority of the TAT scheme and attest to the robustness of our hybrid system.

%\section{Squeezing parameter and its asymptotic behavior in decoherence}\label{sec:4}
\section{Spin squeezing and asymptotic behavior in the presence of dissipation}\label{sec:dissipation}

In this section we shall investigate the squeezing performance of different schemes under decoherence which is more realistic situation. We will extract the optimal squeezing parameter in the time evolution and analyze its scaling behavior with respect to the spin number $N$. We find an asymptotic behavior of the minimum squeezing parameter in fixed system parameters and finally obtain an improved scaling in optimized system parameters.

\subsection{Master equation and link to squeezing schemes}

%%%%%%%%%%%%%%%%%%%%%%%%%%%%%%%%%%%%%%%%%%%%%%%%%%%%%%%%%
%%%%%%%%%%%%%%%%%%%%%%%%%%%%%%%%%%%%%%%%%%%%%%%%%%%%%%%%%
\begin{figure}[t]
%\centering
\includegraphics[width=1\columnwidth]{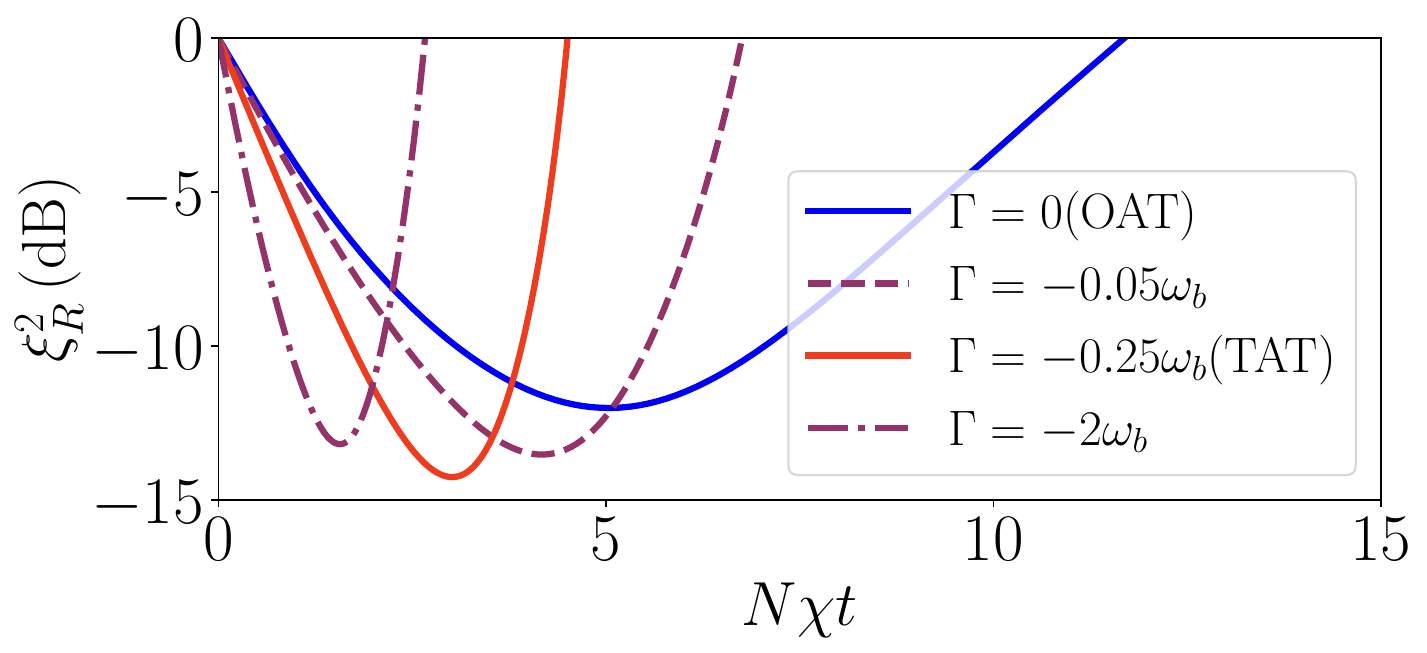}
  \caption{Time evolution of the squeezing parameter [$\xi _R ^2\left( \mathrm{dB} \right) =10\log_{10}\xi _R ^2$] under the four schemes in Eq.\eqref{con:8} in the absence of dissipation, with $N=100$.  The system parameters are chosen as $g/(2\pi) = 1\,\mathrm{kHz}$, $\omega_b/(2\pi) = 1\,\mathrm{GHz}$, and $\tilde{\Omega} = 0$. The blue (red) solid line represents the evolution under the OAT (TAT) scheme, while the purple dashed and dotted-dashed lines correspond to schemes with $\Gamma = -0.05\omega_b$ and $\Gamma = -2\omega_b$, respectively.}
  \label{fig-Squz-time-no-Dissipation}
\end{figure}
%%%%%%%%%%%%%%%%%%%%%%%%%%%%%%%%%%%%%%%%%%%%%%%%%%%%%%%%%
%%%%%%%%%%%%%%%%%%%%%%%%%%%%%%%%%%%%%%%%%%%%%%%%%%%%%%%%%

%%%%%%%%%%%%%%%%%%%%%%%%%%%%%%%%%%%%%%%%%%%%%%%%%%%%%%%%%
%%%%%%%%%%%%%%%%%%%%%%%%%%%%%%%%%%%%%%%%%%%%%%%%%%%%%%%%%
\begin{figure}[t]
%\centering
\includegraphics[width=1\columnwidth]{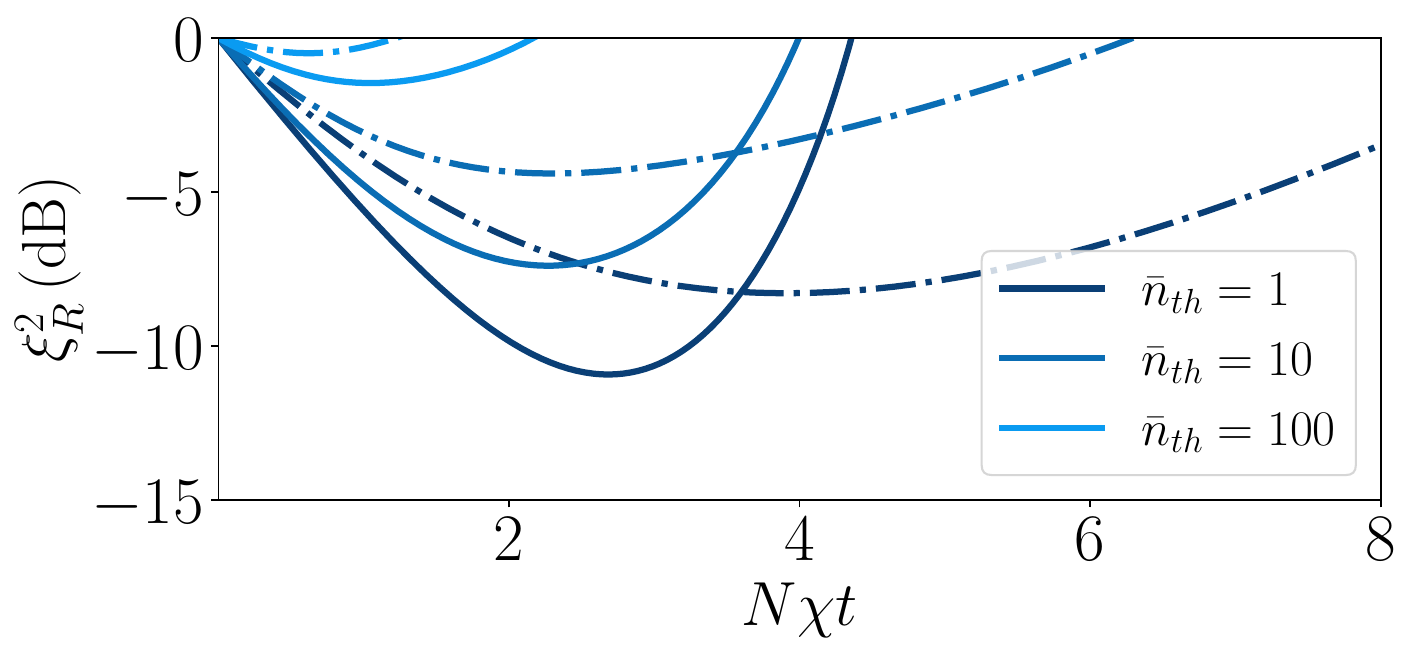}
  \caption{Time evolution under OAT (dotted-dashed line) and TAT (solid line) schemes in the presence of dissipation. The dark blue, blue, and light blue lines indicate the evolutions at different thermal phonon occupations $\bar{n}_{th}= 1$, 10, and 100 for two schemes. Here we set $N=100$, $\omega _r/(2\pi)=200$kHz, $Q_m=10^6$, $g/2\pi = 1\,\mathrm{kHz}$ and a spin dephasing time $T_2=0.01\,\mathrm{s}$.}
  \label{fig-Squz-time-in-Dissipation}
\end{figure}
%%%%%%%%%%%%%%%%%%%%%%%%%%%%%%%%%%%%%%%%%%%%%%%%%%%%%%%%%
%%%%%%%%%%%%%%%%%%%%%%%%%%%%%%%%%%%%%%%%%%%%%%%%%%%%%%%%%

To investigate spin squeezing under decoherence we assume that each two-level system undergoes spin dephasing (with a dephasing rate $T_2^{-1}$), while the MR couples to a thermal bath (characterized by the dissipation rate $\gamma$) and exchanges energy with it. The dissipative dynamics is therefore governed by the master equation (see the derivation in Appendix~\ref{Appendix-Disspation})
\begin{align}\label{con:11}
  \dot{\hat{\rho}}=&-i[\hat{H}_s,\hat{\rho}]+\frac{1}{2T_2}\sum_k{\mathcal{D}}[\hat{\sigma}_{(k)}^{z}]\hat{\rho}\nonumber
  \\
  +&\varGamma _\gamma\left( \bar{n}_{\mathrm{th}}+1 \right)\mathcal{D} [\hat{\mathcal{Z}}] \hat{\rho}+ \varGamma _\gamma\bar{n}_{\mathrm{th}}\mathcal{D}[\hat{\mathcal{Z}}^{\dagger}] \hat{\rho}.
\end{align}
Here, $\mathcal{D}[\hat{A}] \hat{\rho}=\hat{A}\rho \hat{A}^{\dagger}-( \hat{A}^{\dagger}\hat{A}\hat{\rho}+\hat{\rho}\hat{A}^{\dagger}\hat{A}) /2$ denote the Lindblad superoperators, and $\hat{A}$ are the corresponding jump operators. The first term represents the unitary evolution of the system, the second term accounts for the dephasing of individual spins. The last two terms describe the phonon-induced collective spin dissipation, with an effective dissipation rate $\varGamma _{\gamma}=\gamma g^2/\omega _{r}^{2}$. Here, $\gamma =\omega _b/Q_m$ denotes the mechanical damping rate, $\bar{n}_{th}=(e^{\hbar \omega _b/k_BT}-1)^{-1}$ depicts the thermal phonon occupation at temperature $T$, and $Q_m$ is the mechanical quality factor. Notably, the corresponding jump operator for this dissipation process is given by
\begin{align}\label{con:12}
  \hat{\mathcal{Z}}=e^{-2r}\hat{S}_x-ie^{2r}\hat{S}_y,
\end{align}
indicating that the dissipation mechanism is intrinsically linked to the squeezing scheme.

\subsection{Time evolution of the squeezing parameter for
different squeezing schemes in dissipation}

To illustrate the time evolution of the squeezing parameter for
different squeezing schemes in dissipation, we assume the system parameters in orders of $\omega _b/2\pi \sim 1\,\mathrm{GHz}$, $g/2\pi \sim 1\,\mathrm{kHz}$, $Q_m \sim 10^6$~\cite{PhysRevLett.110.156402,chan2011laser,riedinger2016non} and take the dephasing time for individual spin by $T_2 \sim 0.01\,\mathrm{s}$~\cite{bourassa2020entanglement}. For an ensemble with $N = 50$ spins and under several different phonon excitation numbers $\bar{n}_{th}$, we numerically solved the master equation \eqref{con:11} \cite{johansson2012qutip,PhysRevA.98.063815} to obtain the time evolution of the squeezing parameter under both the OAT and TAT schemes, as shown in Fig.~\ref{fig-Squz-time-in-Dissipation}. We see that considerable squeezing effect is still available in the the presence of dissipation, despite the squeezing degree is somewhat weakened relatively to the ideal situation in Fig.~\ref{fig-Squz-time-no-Dissipation}.

The results indicate that, for identical system parameters, the TAT scheme consistently outperforms the OAT scheme. Moreover, as the average excitation number increases, the available squeezing resources in both schemes gradually vanish. A straightforward explanation is that a larger $\bar{n}_{th}$ leads to more vigorous energy exchange between the thermal reservoir and the atomic ensemble, thereby reducing inter-spin coherence during the evolution and driving the ensemble farther away from unitary dynamics.

%%%%%%%%%%%%%%%%%%%%%%%%%%%%%%%%%%%%%%%%%%%%%%%%%%%%%%%%%
%%%%%%%%%%%%%%%%%%%%%%%%%%%%%%%%%%%%%%%%%%%%%%%%%%%%%%%%%
\begin{figure}[t]
\includegraphics[width=1.0\linewidth]{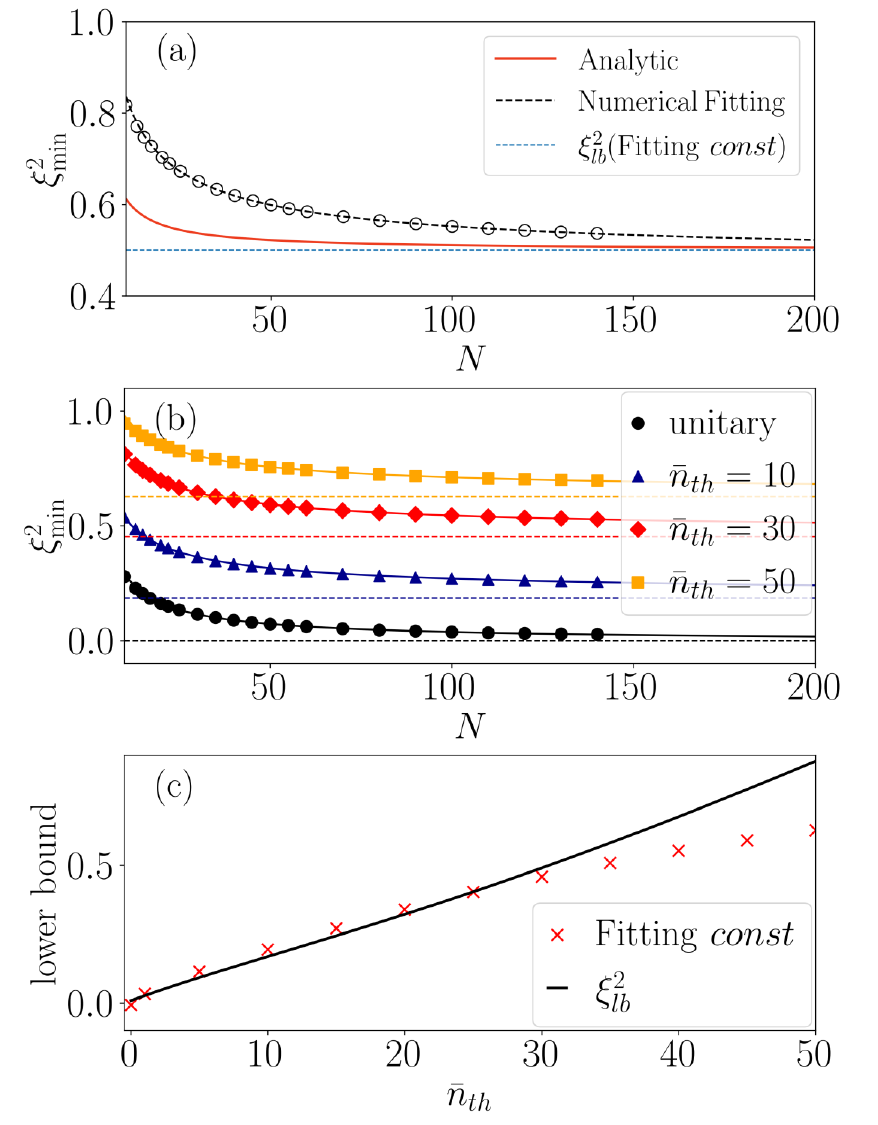}
  \caption{Asymptotic behavior of the squeezing parameter in dissipation.
  (a) Comparison between numerical fitting and analytical results. The red solid line corresponds to the analytical expression given by Eq.\eqref{con:16}, while the black dashed line represents the fitted curve based on Eq.\eqref{con:19}. The blue dashed line indicates the asymptotic squeezing bound obtained from both approaches. Parameters are $\omega_r/(2\pi) = 53$ kHz and $\bar{n}_{th} = 20$.
  (b) Dependence of the squeezing bound on $\bar{n}_{th}$. From bottom to top, the curves correspond to the unitary case and $\bar{n}_{th}= 10, 30, 50$, respectively, with fitting according to Eq.\eqref{con:19}. Here, $\omega_r /(2\pi)= 80$kHz. The remaining parameters in each panel are $g/2\pi =1\mathrm{kHz}, Q_m=10^6, T_2=0.01\mathrm{s}$.
  (c) Lower bound $\xi ^2_{lb}$ versus number of $\bar{n}_{th}$. The black solid line represents the result from Eq.\eqref{con:18}, while the red crosses correspond to the fitting based on Eq.\eqref{con:19}. The parameters used are the same as those in panel (b).
  }
  \label{fig4}
\end{figure}
%%%%%%%%%%%%%%%%%%%%%%%%%%%%%%%%%%%%%%%%%%%%%%%%%%%%%%%%%
%%%%%%%%%%%%%%%%%%%%%%%%%%%%%%%%%%%%%%%%%%%%%%%%%%%%%%%%%

\subsection{Asymptotic behavior of the minimum squeezing parameter in the larger-$N$ limit without parameter optimization}

From the time evolutions of $\xi^2_{R}$ addressed in the last subsection for fixed spin number $N$, we have seen a minimum squeezing parameter $\xi^2_{\rm min}$ can be reached at a certain tim $t_{\rm min}$, which means a maximum squeezing degree available in time evolution.  In this subsection we shall analyze the scaling relation of $\xi^2_{\rm min}$ with respect to the variation of spin number and extract its asymptotic behavior in larger-$N$ limit.

\subsubsection{Dynamical equations for collective spin moments}

By substituting  Eq.~\eqref{con:11} into $\partial _t \langle \hat{A} \rangle =\mathrm{tr}\{\hat{A}\dot{\hat{\rho}}\}$, we derive a set of differential equations about the collective spin moments $\langle \hat{S}_{y}^{2} \rangle$, $ \langle \hat{S}_{z}^{2} \rangle$, and $\langle \hat{S}_yS_z+\hat{S}_zS_y \rangle /2$, which results in an infinite hierarchy of coupled equations. To close the set of equations, we assume that all collective spin correlations beyond second order vanish, allowing factorization of third-order moment as $\langle ABC\rangle \approx \langle A\rangle \langle BC\rangle +\langle B\rangle \langle CA\rangle +\langle AB\rangle \langle C\rangle -2\langle A\rangle \langle B\rangle \langle C\rangle $~\cite{kubo1962generalized}. Since the linearized equations are valid only at short times, we employ the initial conditions corresponding to CSS, given by $\langle \hat{S}_x \rangle =N/2$, $\langle \hat{S}_{x}^{2} \rangle =N^2/4$, and $\langle \hat{S}_y \rangle =\langle \hat{S}_z \rangle =\langle \hat{S}_x\hat{S}_y+\hat{S}_y\hat{S}_x \rangle /2=\langle \hat{S}_x\hat{S}_z+\hat{S}_z\hat{S}_x \rangle /2=0$, these linearized differential equations can be rewritten as
\begin{align}\label{con:13}
\frac{d}{dt}\mathbf{X}=\mathbf{A}\mathbf{X}+\mathbf{M}.
\end{align}
Here, we have defined
\begin{subequations}\label{con:14}
  \begin{align}
    \mathbf{A}&=\left( \begin{matrix}
    	-\left( c+2/T_2 \right)&		c&		\sqrt{2}\chi N\\
	    c&		-5c&		\sqrt{2}\chi N\\
	    \sqrt{2}\chi N/2&		\sqrt{2}\chi N/2&		-\left( 4c+1/T_2 \right)\\
                        \end{matrix} \right)\\
    \mathbf{X}&=\left( \left< S_{y}^{2} \right> , \left< S_{z}^{2} \right> , \left< S_yS_z+S_zS_y \right> /2 \right) ^{\mathrm{T}}\\
    \mathbf{M}&=\left( N/2T_2, cN(N+2), 0 \right) ^{\mathrm{T}}.
  \end{align}
\end{subequations}

At this stage, we have derived the dynamical equations governing the collective spin operator $\mathbf{X}$, which determines the squeezing parameter $\xi _{R}^{2}$. Although this system of equations admits an analytical solution, the resulting expressions are too complex to be presented explicitly here.

\subsubsection{Analytical minimum squeezing parameter $\xi ^2 _{\rm min}$}

To estimate the minimum squeezing parameter $\xi ^2 _{\rm min}$, we assume that the number of two-level atoms $N$ in the system is sufficiently large to satisfy the relation $N\gg N\chi t\gg 1$. Additionally, we posit that all noise sources in the system are weak enough such that the system's evolution is primarily governed by its unitary dynamics. Under these assumptions, approximate expressions for the minimum squeezing parameters and their corresponding squeezing times can be extracted as (see the derivation in Appendix~\ref{Appendix-Opt-Sqz})
\begin{align}\label{con:16}
  \xi _{\rm min}^{2}\simeq M \cdot \left[ \frac{\sqrt{2}c}{\chi}+\frac{1}{N}\frac{\sqrt{2}}{\chi}\left( 2c-\frac{1}{2T_2} \right) \right],
\end{align}
and\begin{align}\label{con:15}
t_{\rm min}\simeq \frac{\Theta}{\sqrt{2}N\chi}.
\end{align}
Here, $\Theta$ and  $M$ are defined as
\begin{subequations}\label{con:17}
  \begin{align}
    \Theta &=\ln \left( 1-\frac{2\chi A_{-}+\frac{\sqrt{2}}{2T_{2}}}{2\chi A_{+}} \right),
     \\
    M &=1+\frac{A_{-}}{e^{2\Theta}A_{+}}-\sqrt{\Theta ^2+\left( 1-\frac{1}{e^{\Theta}} \right) ^2\left( 1+\frac{A_{-}}{e^{\Theta}A_{+}} \right) ^2},
  \end{align}
\end{subequations}
where $A_{\pm}=N\left( 1\pm \sqrt{2}c/\chi \right) /8\pm \sqrt{2}\left( c+1/4T_2 \right) /4\chi $ and parameter $c=\Gamma _\gamma\bigl( \bar{n}_{\mathrm{th}}+1/2 \bigr) $ is associated with the effective spin dissipation induced by mechanical dissipation.

\subsubsection{Asymptotic minimum squeezing parameter in large-$N$ limit}

In the large-$N$ limit, $\Theta$ asymptotically approaches a parameter determined by the system parameter $\chi$ and the dissipation $c$, given by $\Theta \rightarrow \ln  \varepsilon \equiv \ln \left[ 4\:c/\left( \sqrt{2}\chi +2c \right) \right]   $. Meanwhile, in the expression for the optimum squeezing parameter, the ratios $A_{-}/A_{+}\rightarrow 1-\varepsilon $ and $e^{\Theta}\rightarrow \varepsilon$.
This indicates that the squeezing parameter $\xi _{\rm min}^{2}$ cannot decrease indefinitely but instead approaches a lower bound $\xi _{lb}^{2}$ determined by $\varepsilon$, expressed as
\begin{align}\label{con:18}
  \xi _{lb}^{2}= \frac{\varepsilon}{2-\varepsilon}\left[ 1+\frac{(1-\varepsilon )}{\varepsilon ^2}-\sqrt{\ln ^2\varepsilon +\frac{(1-\varepsilon )^2}{\varepsilon ^4}} \right].
\end{align}
In other words, our results indicate that dissipation not only modifies the ideal TAT scaling, but also imposes a lower bound on the achievable squeezing.

To verify this conclusion, we numerically solve Eq.~\eqref{con:11} and fit the dependence of the optimal squeezing parameter $\xi_{\rm min}^2$ on $N$ using a power-law function with an additive constant $const$, expressed as
\begin{align}\label{con:19}
\xi _{\rm min}^{2}=aN^b+const.
\end{align}
The scaling curves obtained from numerical fitting and Eq.~\eqref{con:16} are plotted in Fig.~\ref{fig4}(a), where one can unambiguously observe that the $\xi_{ opt}^2$ from both methods converges asymptotically, and their lower bounds coincide with a satisfactory precision.

To further cement our conclusions, Fig.~\ref{fig4}\hyperref[fig4]{(b)} shows the $\xi_{\rm min}^2$ scaling under different dissipation strengths-parametrized by the phonon excitation $\bar{n}_{th}$. Strikingly, as dissipation is reduced, the squeezing bound steadily diminishes, ultimately collapsing to zero in the ideal situation. This unequivocally demonstrates that dissipative processes indeed impose a finite lower bound on achievable spin squeezing, regardless of how large $N$ becomes. The squeezing parameter \eqref{con:10} as the ratio between the phase sensitivity of an arbitrary state and CSS~\cite{PhysRevA.46.R6797,PhysRevA.50.67}, is given by $\xi _{R}^{2}=(\Delta \phi )^2/(\Delta \phi )_{\mathrm{CSS}}^{2}$. $\xi _{\rm min}^{2}$ approaching $\xi _{lb}^{2}$ implies that the metrological precision of the SSS asymptotically returns to the $N^{-1/2}$ of the CSS. This indicates that under the influence of dephasing of individual spins and phonon-induced collective spin dissipation, the metrological gain offered by spin squeezing saturates as the spin number $N$ increases.

Furthermore, the analytical squeezing limit obtained from the linearized method [Eq.\eqref{con:18}] exhibits an excellent agreement with the numerical fitting in the weak dissipation regime, as shown in Fig.~\ref{fig4}\hyperref[fig4]{(c)}. As the dissipation increases, deviations appear due to the breakdown of the weak-coupling approximation. This behavior indicates that the approximations adopted in our linearized analysis capture the essential physics of the dissipation process, thereby validating the method.

\subsection{Improved scaling relation for optimal squeezing in parameter optimization}

As a final result, we present an improved scaling relation from the optimal squeezing parameter $\xi_{\rm opt}^{2}$ by optimization over the parameter $\omega _r$. Although at a fixed system parameter $\omega _r$ a deeper spin squeezing in large-$N$ cases seems to be somewhat hindered by the asymptotic limit as discussed in the last subsection, such a limitation may be broken in optimization of $\omega _r$. Indeed, such a parameter optimization releases more potential of quantum resource for the spin squeezing, leading to an improved scaling relation for the squeezing parameter.

\begin{figure}[t]
  \includegraphics[width=1.0\linewidth]{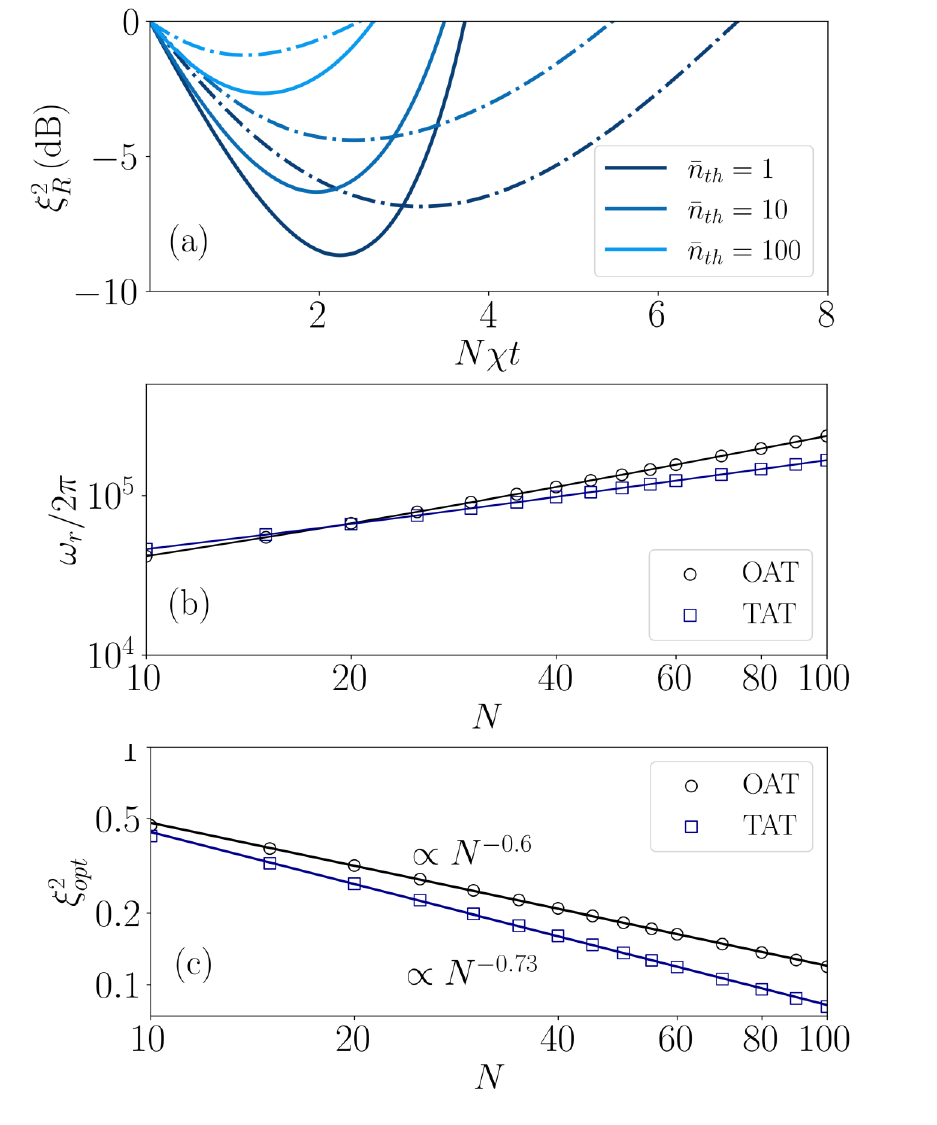}
  \caption{Scaling of the optimal squeezing with respect to the spin number $N$ for the OAT and TAT schemes in dissipation.
  (a) Time evolutions of the squeezing parameter $\xi ^2_{R}$ for a certain spin number $N$ at optimized parameter of $\omega _r$. The other parameters are the same as Fig.~\ref{fig-Squz-time-in-Dissipation}.
  (b) Dependence of the optimized $\omega _r$ over the spin number.
  (c) The scaling relation of the optimal squeezing parameter $\xi ^2_{\rm opt}$ with respect to $N$. For each $N$, $\omega_r$ is optimized to achieve the maximized squeezing degree (minimum of $\xi ^2_{\rm min}$).
  In (b) and (c), $\bar{n}_{th}=1$, $g/2\pi =1\mathrm{kHz}, Q_m=10^6, T_2=0.01\mathrm{s}$. The circles and squares in (a) and (b) represent the OAT and TAT numerical results respectively, while the black and blue solid lines are the corresponding fitted scaling.
  }
  \label{Fig-Scaling-optimal-squz}
\end{figure}

In Fig.\ref{Fig-Scaling-optimal-squz}(a) we illustrate some time evolutions at optimized $\omega _r$ for a certain spin number, which yield lower minimum values $\xi^2_{\rm min}$ in time evolution than the unoptimized cases in Fig.~\ref{fig-Squz-time-in-Dissipation}. The dependence of the optimized $\omega _r$ is given in Fig.\ref{Fig-Scaling-optimal-squz}(b) for both OAT (circles) and TAT (squares).

In the ideal situation, the OAT squeezing scheme scales as $\xi_{R}^{2} \propto N^{-2/3}$, while the TAT scheme follows $\xi_{R}^{2} \propto N^{-1}$~\cite{PhysRevA.47.5138}. However, the presence of dissipation degrades these ideal scaling relations. By numerically solving Eq.~\eqref{con:11} with $\omega_{r}$ optimized for each $N$ and employing the same parameters in the Fig.~\ref{fig4}(b) with $\bar{n}_{th}=1$, in Fig.~\ref{Fig-Scaling-optimal-squz}(c) we obtain the fitted scaling relations for the two schemes when decoherence effects are incorporated:
\begin{eqnarray}
\xi_{R}^{2} \propto N^{-0.6}  \quad \text{(OAT)},
\\
\xi_{R}^{2} \propto N^{-0.73} \quad \text{(TAT)}.
\end{eqnarray}
These scaling relations turn out to surpass many existing schemes in the literature, where $\xi_{R}^{2} \propto N^{-0.5}$~\cite{PhysRevLett.110.156402,PhysRevLett.125.203601}, $N^{-0.54}$~\cite{BaiSY2021PRL} and $N^{-0.64}$~\cite{ren2024heisenberg}, demonstrating that our system provides a valuable reference for the experimental realization.

%%%%%%%%%%%%%%%%%%%%%%%%%%%%%%%%%%%%%%%%%%%%%%%%%%%%%%%%%
%%%%%%%%%%%%%%%%%%%%%%%%%%%%%%%%%%%%%%%%%%%%%%%%%%%%%%%%%
%%%%%%%%%%%%%%%%%%%%%%%%%%%%%%%%%%%%%%%%%%%%%%%%%%%%%%%%%
%%%%%%%%%%%%%%%%%%%%%%%%%%%%%%%%%%%%%%%%%%%%%%%%%%%%%%%%%

\subsection{Shortened preparation time for optimal squeezing in parameter optimization}

\begin{figure}[t]
  \includegraphics[width=1.0\linewidth]{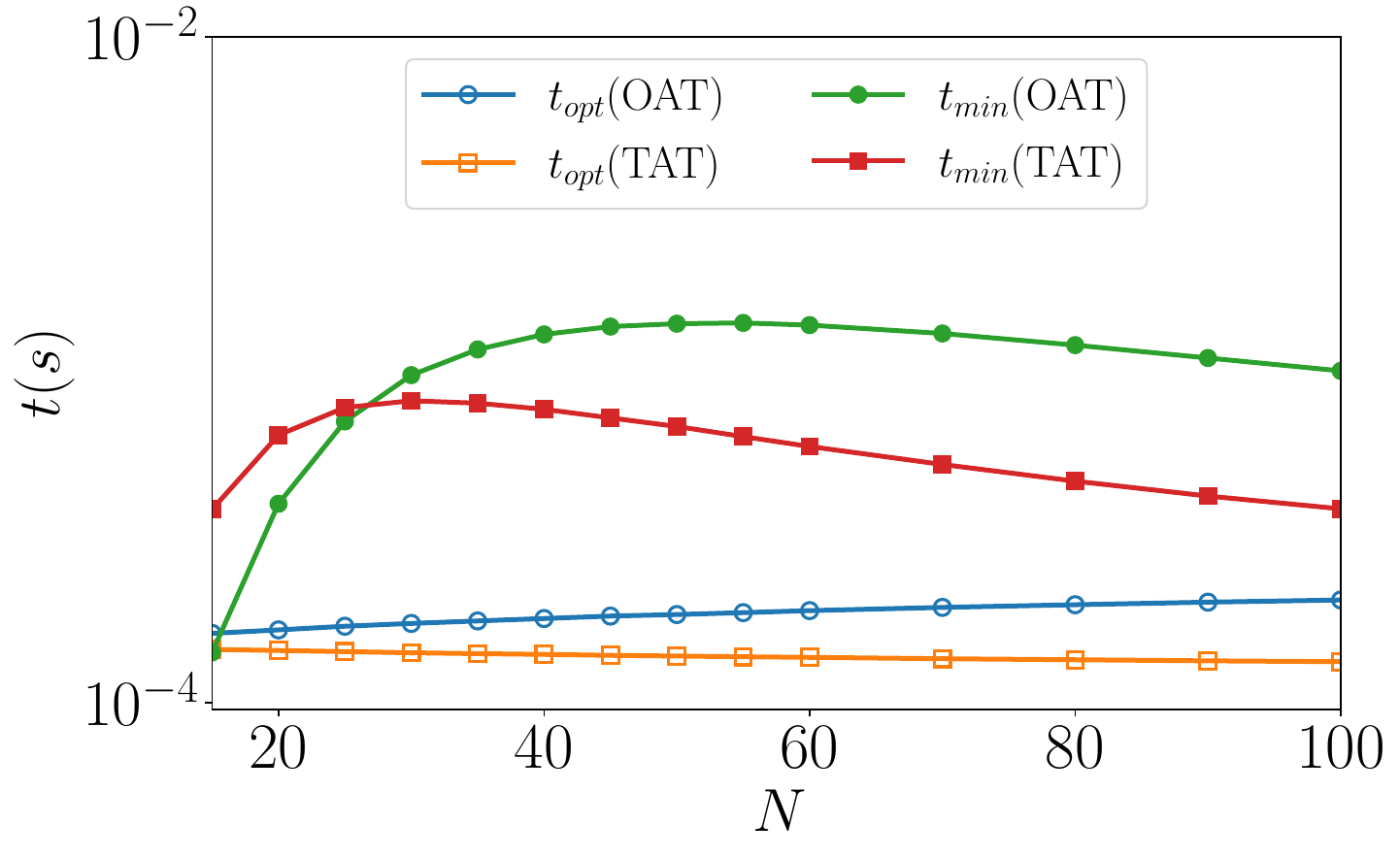}
  \caption{Shortened preparing time for optimal squeezing $\xi ^2 _{\rm opt}$ in optimized $\omega_r$ ($t_{\rm opt}$, empty symbols) compared with the preparing time for $\xi ^2 _{\rm min}$ in unoptimized $\omega_r$ ($t_{\rm min}$, filled symbols) in OAT (round) and TAT (square). Here the unoptimized case is illustrated by $\omega_r/(2\pi)=800$kHz which is comparable to the optimized ones, the other parameters for $t_{\rm opt}$ and $t_{\rm min}$ are the same as in Fig.~\ref{Fig-Scaling-optimal-squz}(c) and Fig.~\ref{fig-Squz-time-in-Dissipation} with $\bar{n}_{th}=1$, respectively.
  }
  \label{Fig-time-optimzation}
\end{figure}

Besides the measurement precision another major concern in quantum metrology is reduction of the preparation time for probe state~\cite{
Ying2022-Metrology,Ying-g2hz-QFI-2024,Ying-g1g2hz-QFI-2025,Ying2025g2A4,Ying2025TwoPhotonStark,Gietka2022-ProbeTime}. Needless to say, a shorter time of preparation for probe state would be more favorable in practical applications. Here for the phonon-induced spin squeezing we find that, apart from the smaller value of $\xi ^2 _{\rm opt}$ in optimized $\omega_r$ than $\xi ^2 _{\rm min}$ in unoptimized $\omega_r$, the preparation time $t_{\rm opt}$ to reach $\xi ^2 _{\rm opt}$ in the optimized case is also shortened than the preparation time $t_{\rm min}$ to reach $\xi ^2 _{\rm min}$ in the unoptimized cases. This contrast can be clearly seen in the illustrations of Fig.~\ref{Fig-time-optimzation} where $t_{\rm opt}$ is considerably reduced in the entire regime of spin number $N$ and for both OAT and TAT.

Thus, with parameter optimization in our hybrid spin-optomechanical system, we have several merits of decoherence reduction for the phonon-induced spin squeezing: (i) Diverse
spin-squeezing protocols, including OAT, TAT and two-axis squeezing with different weightings; (ii) Larger squeezing degree in optimal squeezing and improved scaling relation with respect to the spin number outperforming the existing schemes; (iii) Shortened preparation time for optimal squeezing. These advantages should be favorable for potential applications of our schemes in spin squeezing and quantum metrology.

%%%%%%%%%%%%%%%%%%%%%%%%%%%%%%%%%%%%%%%%%%%%%%%%%%%%%%%%%
%%%%%%%%%%%%%%%%%%%%%%%%%%%%%%%%%%%%%%%%%%%%%%%%%%%%%%%%%
%%%%%%%%%%%%%%%%%%%%%%%%%%%%%%%%%%%%%%%%%%%%%%%%%%%%%%%%%
%%%%%%%%%%%%%%%%%%%%%%%%%%%%%%%%%%%%%%%%%%%%%%%%%%%%%%%%%

\section{CONCLUSION}\label{sec:CONCLUSION}

In this work, we have demonstrated versatile Heisenberg-limited spin squeezing schemes in a hybrid optomechanical system comprising an atomic ensemble, where the collective spin is coupled to mechanical resonator via a TC interaction.

By linearizing the optomechanical coupling and adiabatically eliminating the optical and mechanical degrees of freedom in the regime where the frequencies of atomic ensemble and optical mode are far detuned from the phonon frequency, we engineer an effective Bogoliubov-type collective spin-spin interaction that enables a robust and tunable protocol capable of realizing different dynamics in OAT, TAT, and two-axis schemes with various weighings. The versatile spin squeezing schemes are illustrated by the Husimi Q function and more quantitatively demonstrated by the squeezing parameter $\xi_{\rm min}^2$ which characterizes the squeezing degree.

Besides the ideal situation in the absence of dissipation, we have also investigated the spin squeezing degree in the presence of single-spin dephasing and mechanical dissipation. We have found that the minimum squeezing parameter does not indefinitely decrease to zero in the large-$N$ limit, but instead asymptotically approaches a finite value determined jointly by the system parameters $\chi$ and the equivalent dissipation rates $c$. In terms of a linearized approach we have further derived an analytical expression for this asymptotic bound in the large-$N$ limit under weak dissipation, which is confirmed by numerical results.
Our results reveal that the metrological precision achievable via spin squeezing asymptotically degrades to the standard quantum limit with increasing particle number $N$ under Markovian decoherence process, reflecting a loss of quantum advantage. Nevertheless, we also consider the optimization of system parameters which can release more potential of quantum resources and compensate for such loss. The optimization finally leads to an optimal squeezing, which achieves improved scaling relations of the squeezing parameter with respect to the spin number $N$, being cable of surpassing existing schemes in
the literature. Moreover, the optimization also leads to a considerable reduction of the preparation time for the optimal squeezing.

To wrap up, our results enable more robust and adjustable spin squeezing schemes. The obtained asymptotic behavior for optimal squeezing parameter in large-$N$ limit may provide some unconventional insights into the realistic performance of spin-squeezed states and offer some useful guidelines for experimental realizations in quantum open systems. On the other hand, our scheme not only yields improved scaling relations for the degree of spin squeezing, but also shortens the time to reach the optimal squeezing.

\begin{acknowledgments}
This work was supported by the National Natural Science Foundation of China
(Grants No. 12474358, No. 11974151, and No. 12247101).
\end{acknowledgments}

\appendix

\section{Linearization of Optomechanical Interaction}\label{Appendix-Linearize-OM}

Under the influence of environmental noise, the dynamics of Hamiltonian \eqref{con:2} is governed by the quantum Langevin equation:
\begin{subequations}\label{con:A1}
  \begin{align}
    \dot{\hat{a}}&=-\left( i\Delta _a+\frac{\kappa _a}{2} \right) \hat{a}-ig_0\hat{a}\left( \hat{b}+\hat{b}^{\dagger} \right) -i\Omega _a+\sqrt{\kappa _a}\hat{a}_{in},\\
    \dot{\hat{b}}&=-\left( i\omega _b+\frac{\kappa _b}{2} \right) \hat{b}-ig_0\hat{a}^{\dagger}\hat{a}-ig\hat{S}_-+\sqrt{\kappa _b}\hat{b}_{in},
  \end{align}
\end{subequations}
where $\hat{a}_{in}$ and $\hat{b}_{in}$ denote the environmental noise for the cavity field and the MR, with $\kappa _a$ and $\kappa _b$ representing the corresponding dissipation rates. Under strong laser driving, the operators can be decomposed as $\hat{a} = \alpha + \delta\hat{a}$, $\hat{b} =\beta + \delta\hat{b}$. By replacing $(\delta\hat{a}, \delta\hat{b})$ with $(\hat{a},\hat{b})$ and neglecting higher-order fluctuation terms, the dynamics of the fluctuation operators are determined by
\begin{subequations}\label{con:A2}
  \begin{align}
    \dot{\hat{a}}&=-\left[ i(\Delta _a+2g_0\beta )+\frac{\kappa _a}{2} \right] \hat{a}-ig_0\alpha \left( \hat{b}+\hat{b}^{\dagger} \right) +\sqrt{\kappa _a}\hat{a}_{in}\\
    \dot{\hat{b}}&=-\left( i\omega _b+\frac{\kappa _b}{2} \right) \hat{b}-ig_0 \alpha \left( \hat{a}+\hat{a}^{\dagger} \right) -igS_-+\sqrt{\kappa _b}\hat{b}_{in}.
  \end{align}
\end{subequations}
By comparing the linearized Heisenberg-Langevin equation \eqref{con:A2} with the standard form, the corresponding effective Hamiltonian can be expressed as follows
\begin{align}
  \hat{H}_L&=\left( \Delta _a+2g_0\beta \right) \hat{a}^{\dagger}\hat{a}+\omega _b\hat{b}^{\dagger}\hat{b}+\Omega \hat{S}_z, \nonumber
  \\
  &+g_0\alpha \left( \hat{a}+\hat{a}^{\dagger} \right) \left( \hat{b}+\hat{b}^{\dagger} \right) +g\left( \hat{b}\hat{S}_++\hat{b}^{\dagger}\hat{S}_- \right),
\end{align}
which gives the linearized effective Hamiltonian in Eq.~\eqref{con:3} in the main text.

\section{The derivation of the $\hat{H}_{eff}$ in Eq.\eqref{con:5}}\label{Appendix-Heff}

We decompose Hamiltonian \eqref{con:3} into a free part and an interaction part:
\begin{subequations}\label{con:B1}
 \begin{align}
	\hat{H}_0&=\Delta \hat{a}^{\dagger}\hat{a}+\omega _b\hat{b}^{\dagger}\hat{b}+\Omega \hat{S}_z+g\left( \hat{b}\hat{S}_++\hat{b}^{\dagger}\hat{S}_-\right),\\
	\hat{H}_I&=G\left( \hat{a}+\hat{a}^{\dagger} \right) \left( \hat{b}+\hat{b}^{\dagger} \right).
 \end{align}
\end{subequations}
In the large-detuning regime between the optical mode $\hat{a}$ and the phonon mode $\hat{b}$, i.e., $G\ll |\Delta -\omega _b|$, the interaction strength $G$ is relatively weak compared to the system, allowing us to treat $\hat{H}_I$ as a perturbation. Consequently, we employ the Fr\"{o}hlich-Nakajima transformation~\cite{PhysRev.79.845,nakajima1955perturbation}, $U = e^V$, to obtain an effective Hamiltonian for the system. Here, $V$ takes the form
\begin{align}\label{con:B2}
  \hat{V}=\mu \left(\hat{a}\hat{b}-\hat{a}^{\dagger}\hat{b}^{\dagger}\right) +\nu \left(\hat{a}\hat{b}^{\dagger}-\hat{a}^{\dagger}\hat{b}\right).
\end{align}
By appropriately choosing $\mu$ and $\nu$, we ensure that $\hat{H}_I+[\hat{H}_0,\hat{V}]=0$, leading to the transformed Hamiltonian expressed as
\begin{align}\label{con:B3}
\hat{H}_{eff}=\hat{U}^{\dagger}\hat{H}_L\hat{U}=\hat{H}_0+\frac{1}{2}[\hat{H}_I,\hat{V}]+\frac{1}{3}[[\hat{H}_I,\hat{V}],\hat{V}]+\cdots .
\end{align}
Since both $\mu =G/(\Delta+\omega_{b})$ and $\nu =G/(\Delta-\omega_{b})$ are small in the large-detuning regime, we retain only the first two terms of the effective Hamiltonian \eqref{con:B3}, which can be written as:
\begin{align}\label{con:B4}
\hat{H}_{eff}&\approx\hat{H}_0+\frac{1}{2}[\hat{H}_I,\hat{V}]\nonumber
\\
&=\omega _b\hat{b}^{\dagger}\hat{b}+\Omega \hat{S}_z+g\left( \hat{b}\hat{S}_++\hat{b}^{\dagger}\hat{S}_- \right)-\varGamma\left( \hat{b}+\hat{b}^{\dagger} \right) ^2.
\end{align}
Here we have neglected the free energy of the optical field, leading to the effective Hamiltonian given in Eq.\eqref{con:5} in the main text.

\section{Phonon-induced collective spin dissipation}\label{Appendix-Disspation}

The original master equation of our hybrid system is given by
\begin{align}\label{con:C1}
\dot{\hat{\rho}}&=-i[\hat{H}_L,\hat{\rho}]+\frac{1}{2T_2}\sum_k{\left[ \hat{\sigma}_{(k)}^{z}\hat{\rho}\hat{\sigma}_{(k)}^{z}-\hat{\rho} \right]}\nonumber
\\
&+\gamma \left( \bar{n}_{\mathrm{th}}+1 \right) \left[ \hat{b}\hat{\rho}\hat{b}^{\dagger}-\frac{1}{2}\left( \hat{b}^{\dagger}\hat{b}\hat{\rho}+\hat{\rho}\hat{b}^{\dagger}\hat{b} \right) \right] \nonumber
\\
&+\gamma \bar{n}_{\mathrm{th}}\left[ \hat{b}^{\dagger}\hat{\rho}\hat{b}-\frac{1}{2}\left( \hat{b}\hat{b}^{\dagger}\hat{\rho}+\hat{\rho}\hat{b}\hat{b}^{\dagger} \right) \right]
\end{align}
where $\hat{H}_L$ corresponds to Eq.\eqref{con:3} in the main text. Since we apply a series of unitary transformations
\begin{align}\label{con:C2}
\hat{\mathcal{O}}=e^{\hat{V}}\hat{U}(r)e^{\hat{R}}
\end{align}
to simplify Eq.\eqref{con:3} into the effective Hamiltonian Eq.\eqref{con:8}, the jump operators associated with MR dissipation in Eq.\eqref{con:C1} should also be transformed into the intermediate representation, expressed as
\begin{subequations}\label{con:C3}
\begin{align}
    \hat{\mathcal{O}}^{\dagger}\hat{b}\hat{\mathcal{O}}&\approx \left( \hat{b}\cosh r-\hat{b}^{\dagger}\sinh r \right) +\left( \nu \hat{a}-\mu \hat{a}^{\dagger} \right)\nonumber\\
    &-\frac{g}{\omega _r}\left( \cosh r\hat{\varXi}-\sinh r\hat{\varXi}^{\dagger} \right),\\
    \hat{\mathcal{O}}^{\dagger}\hat{b}^{\dagger}\hat{\mathcal{O}}&\approx \left( \hat{b}^{\dagger}\cosh r-\hat{b}\sinh r \right) +\left( \nu \hat{a}^{\dagger}-\mu \hat{a} \right)\nonumber\\
    &-\frac{g}{\omega _r}\left( \cosh r\hat{\varXi}^{\dagger}-\sinh r\hat{\varXi} \right) .
\end{align}
\end{subequations}
Substituting Eq.\eqref{con:C3} into Eq.\eqref{con:C1} and using the relation
\begin{align}\label{con:C4}
&\mathcal{D} \left[ \hat{A}+\hat{B}+\hat{C} \right] \hat{\rho}\nonumber
\\
=&\mathcal{D} \left[ \hat{A} \right] \hat{\rho}+\mathcal{D} \left[ \hat{B} \right] \hat{\rho}+\mathcal{D} \left[ \hat{C} \right] \hat{\rho}\nonumber
\\
+&\mathcal{D} \left[ \hat{A},\hat{B} \right] \hat{\rho}+\mathcal{D} \left[ \hat{B},\hat{C} \right] \hat{\rho}+\mathcal{D} \left[ \hat{C},\hat{A} \right] \hat{\rho}\nonumber
\\
+&\mathcal{D} \left[ \hat{B},\hat{A} \right] \hat{\rho}+\mathcal{D} \left[ \hat{C},\hat{B} \right] \hat{\rho}+\mathcal{D} \left[ \hat{A},\hat{C} \right] \hat{\rho},
\end{align}
where $\mathcal{D} \left[ \hat{A},\hat{B} \right] \hat{\rho}=\hat{A}\hat{\rho}\hat{B}^{\dagger}-\frac{1}{2}\left( \hat{A}^{\dagger}\hat{B}\hat{\rho}+\hat{\rho}\hat{A}^{\dagger}\hat{B} \right)$, we find that the system exhibits phonon dissipation, phonon-induced dissipation of the optical field and the collective spin, as well as their cross-coupling terms. Since our interest is confined to the dissipation of the atomic ensemble, we retain only the spin dissipation terms.

Within Eq.\eqref{con:C3}, the phonon-induced collective spin dissipation operator can be written as
\begin{align}\label{con:C5}
-\frac{g}{\omega _r}\left( \cosh r\hat{\Sigma}-\sinh r\hat{\Sigma}^{\dagger} \right) =-\frac{g}{\omega _r}\left( e^{-2r}\hat{S}_x-ie^{2r}\hat{S}_y \right),
\end{align}
so that the corresponding dissipative term becomes
\begin{align}\label{con:C6}
 \gamma (\bar{n}_{\mathrm{th}}+1)\mathcal{D} [-\frac{g}{\omega _r}\left( e^{-2r}\hat{S}_x-ie^{2r}\hat{S}_y \right) ]\hat{\rho} \nonumber \\
 =\frac{\gamma g^2}{\omega _{r}^{2}}(\bar{n}_{\mathrm{th}}+1)\mathcal{D} [\hat{\mathcal{Z}}]\hat{\rho}.
\end{align}
The other dissipation term is treated in a similar manner and ultimately we obtain Eq.\eqref{con:11} in the main text.

\section{Analytical estimation of the maximum squeezing}\label{Appendix-Opt-Sqz}

We first solve the linearized equations Eq.\eqref{con:13} in the large-$N$ limit, where the evolution is predominantly governed by the unitary dynamics. For computational convenience, we arrange the solution of $\mathbf{X}$ as
\begin{subequations}\label{con:D1}
\begin{align}
    \left< S_{y}^{2}-S_{z}^{2} \right> &=-Pt,
    \\
    \left< S_{y}^{2}+S_{z}^{2} \right> &=2A_+e^{\sqrt{2}N\chi t}+2A_-e^{-\sqrt{2}N\chi t},
    \\
    \left<C_{yz}\right> &=A_+\left( e^{\sqrt{2}N\chi t}-1 \right) -A_-\left( e^{-\sqrt{2}N\chi t}-1 \right) ,
\end{align}
\end{subequations}
Here, $A_+$, $A_-$, and $P$ denote
\begin{subequations}\label{con:D2}
\begin{align}
A_+&=\frac{N}{8}\left( 1+\frac{\sqrt{2}c}{\chi} \right) +\frac{\sqrt{2}}{4\chi}\left( c+\frac{1}{4T_2} \right),
\\
A_-&=\frac{N}{8}\left( 1-\frac{\sqrt{2}c}{\chi} \right) -\frac{\sqrt{2}}{4\chi}\left( c+\frac{1}{4T_2} \right),
\\
P&=N\left[ cN+\left( 2c-\frac{1}{2T_2} \right) \right].
\end{align}
\end{subequations}
In the large-$N$ limit, because the initial state is polarized along the $x$ axis and the polarization direction remains nearly unchanged at short times during the evolution, we have $| \left< S \right>|^2=| \left< S_x \right>|^2=N^2/4$. From Eq.\eqref{con:11}, it follows that
\begin{align}\label{con:D3}
\frac{d  \xi _{R}^{2}  }{dt}=\frac{4}{N}\frac{d\left<(\Delta S)^2 \right> _{\perp}}{dt}.
\end{align}
By substituting Eq.\eqref{con:D1} into Eq.\eqref{con:D3} and assuming that $N\chi t \gg 1$, the solution to Eq.\eqref{con:D3} is given by:
\begin{align}\label{con:D4}
t_{\rm min}=\frac{1}{\sqrt{2}\chi N}\ln\left[1-\frac{2\chi A_-+\frac{\sqrt{2}}{2T_2}}{2\chi A_+}\right].
\end{align}
Inserting this result into Eq.\eqref{con:10} yields the optimal squeezing parameter as
\begin{align}\label{con:D5}
  \xi _{\rm min}^{2}&=\left[ \frac{\sqrt{2}c}{\chi}+\frac{1}{N}\left( \frac{2\sqrt{2}c}{\chi}-\frac{\sqrt{2}}{2\chi T_2} \right) \right] \nonumber
  \\
  &\times \left[ 1+\frac{A_-}{e^{2\Theta}A_+}-\sqrt{\Theta ^2+\left( 1-\frac{1}{e^{\Theta}} \right) ^2\left( 1+\frac{A_-}{A_+e^{\Theta}} \right) ^2} \right].
\end{align}
Note that this result is derived under the condition $N \gg N\chi t \gg 1$. The factor following the optimal squeezing parameter $\xi _{\rm min}^{2}$ corresponds to $M$ in  Eq.~\eqref{con:17}. In the limit $N\rightarrow\infty$, it is straightforward to show that
\begin{align}\label{con:D6}
\xi _{\rm min}^{2}\rightarrow\frac{\varepsilon}{2-\varepsilon}\left[ 1+\frac{(1-\varepsilon )}{\varepsilon ^2}-\sqrt{\ln ^2\varepsilon +\frac{(1-\varepsilon )^2}{\varepsilon ^4}} \right],
\end{align}
where $\varepsilon =4c/(\sqrt{2}\chi +2c)$.

\bibliography{Refs-Spin-Squeezing-2axis}

\end{document}